\documentclass{aastex631}

\newcommand{\source}[1]{\textsuperscript{\textcolor{blue}{[citation needed]}}}

\newcommand{\densunitSI}{kg\,m$^{-3}$}
\newcommand{\kms}{km\,s$^{-1}$}
\newcommand{\ms}{m\,s$^{-1}$}

\newcommand{\fireballTimeZero}{2020-06-19T20:05:07.800}
\newcommand{\codename}{DN200619\_01}
\newcommand{\meteorite}{\textit{Madura Cave}}

\shorttitle{Madura Cave}
\shortauthors{Devillepoix et al.}

\graphicspath{{./}{figs/}}

\begin{document}
\title{Trajectory, recovery, and orbital history of the Madura Cave meteorite}

\newcommand{\curtin}{School of Earth and Planetary Sciences, Curtin University, Perth WA 6845, Australia}

\correspondingauthor{Hadrien A. R. Devillepoix}
\email{hadrien.devillepoix@curtin.edu.au}

\author[0000-0001-9226-1870]{Hadrien A. R. Devillepoix}
\affiliation{\curtin}

\author[0000-0003-2702-673X]{Eleanor K. Sansom}
\affiliation{\curtin}

\author[0000-0003-4766-2098]{Patrick Shober}
\affiliation{\curtin}

\author[0000-0002-8914-3264]{Seamus L. Anderson}
\affiliation{\curtin}

\author[0000-0002-8240-4150]{Martin C. Towner}
\affiliation{\curtin}

\author[0000-0002-5391-3001]{Anthony Lagain}
\affiliation{\curtin}

\author[0000-0003-2193-0867]{Martin Cup\'ak}
\affiliation{\curtin}

\author[0000-0002-4681-7898]{Philip A. Bland}
\affiliation{\curtin}

\author[0000-0002-5864-105X]{Robert M. Howie}
\affiliation{\curtin}

\author[0000-0002-0363-0927]{Trent Jansen-Sturgeon}
\affiliation{\curtin}

\author[0000-0002-8646-0635]{Benjamin A. D. Hartig}
\affiliation{\curtin}

\author[0000-0001-5772-338X]{Marcin Sokolowski}
\affiliation{International Centre for Radio Astronomy Research, Curtin University, Bentley, WA, 6102, Australia}

\author{Gretchen Benedix}
\affiliation{\curtin}

\author{Lucy Forman}
\affiliation{\curtin}

\begin{abstract}
On the 19th June 2020 at 20:05:07 UTC, a fireball lasting $5.5\,s$ was observed above Western Australia by three Desert Fireball Network observatories.
The meteoroid entered the atmosphere with a speed of 
$14.00 \pm 0.17$\kms{} and followed a $58$\degr\ slope trajectory from a height of 75 km down to 18.6 km.
Despite the poor angle of triangulated planes between observatories (29\degr) and the large distance from the observatories, a well constrained kilo-size main mass was predicted to have fallen just South of Madura in Western Australia.
However, the search area was predicted to be large due to the trajectory uncertainties.
Fortunately, the rock was rapidly recovered along the access track during a reconnaissance trip.
The 1.072 kg meteorite called \meteorite{} was classified as an L5 ordinary chondrite.
The calculated orbit is of Aten type (mostly contained within the Earth's orbit), the second time only a meteorite is observed on such an orbit after Bunburra Rockhole.
Dynamical modelling shows that \meteorite{} has been in near-Earth space for a very long time. The NEO dynamical lifetime for the progenitor meteoroid is predicted to be $\sim$\,87\,Myr.
This peculiar orbit also points to a delivery from the main asteroid belt via the $\nu6$ resonance, and therefore an origin in the inner belt.
This result contributes to drawing a picture for the existence of a present-day L chondrite parent body in the inner belt.
\end{abstract}

\section{Introduction} \label{sec:intro}

About half of the meteorites with orbits recovered so far (see \citet{2015aste.book..257B} for a review) have semi-major axes that still identifies the source resonance ($a>2$\, AU) \citep{2020IAUGA..30....9J}.
The other half, on evolved orbits, have dynamically detached themselves from the belt, via close encounters with the inner planets.
In this group, some meteoroids have evolved so much that their aphelion distance (point in the orbit farthest from the Sun) is also out of the belt.
The lessened influence of planetary perturbations from the gas giants significantly lengthens their dynamical lifetimes in near-Earth space.
More rarely, they evolve sufficiently to go onto Aten orbits ($a<1$\,AU).
The objects that follow this evolutionary pathway typically originate in the inner main belt, via the $\nu6$ resonance \citep{2018Icar..311..271G}.
The dominant asteroid types in the inner belt: Vesta family members (V-type) and Flora members (S-type) provide a significant number of HED achondrites and LL chondrites, respectively \citep{2008Natur.454..858V}.

Interestingly, the only meteorite until this present work with a measured Aten orbit is Bunburra Rockhole, an HED but likely not connected to the Vesta clan \citep{2009Sci...325.1525B}.
Bunburra Rockhole likely came from the inner belt, and evolved via the $\nu6$ resonance into the inner planet region.

L chondrites represent a sizable fraction of meteorite falls (33\%), but the search for the parent region and dynamical evolution mechanism of L chondrites is still an ongoing effort.
It was once suggested that shocked L chondrites (2/3\textsuperscript{rd} of L falls) come from the Gefion collisional family via the 5:2 mean-motion resonance with Jupiter \citep{2009Icar..200..698N}, an explanation in principle fitting nicely with fossil L chondrite meteorites found in an $\simeq467$ Ma old geological layer \citep{2001E&PSL.194....1S}.
A major collision event near a powerful resonance transport route is indeed required to quickly fling large amount of debris into near-Earth space.
But, although L chondrites with a $\sim$470\,Ma K-Ar resetting age are still being recovered, it is not clear that these are still transported via the 5:2 resonance today.
It is also not clear the Gefion family forming event is responsible for this  $\sim$470\,Ma K-Ar resetting event in the shocked L chondrites, as \citet{2018MNRAS.476..630M} have shown that reflectance spectra of Gefion family members do not match the mineralogy of L chondrites, and because the Gefion family is likely to be much older \citep{2015Icar..257..275S}.
So whether there exists a collisional family of L chondrite asteroids feeding Near-Earth space is still an open question.
In any case, the existence of a large collisional family is not a necessary condition to produce meteorites.
Meteoroids and small asteroids can be ejected by small impacts, and, thanks to their high Yarkovsky mobility, can access multiple transport routes to near-Earth space.
This is shown by the source region analysis of \citet{2018Icar..311..271G} on shock-darkened L chondrites: L6 Novato \citep{2014M&PS...49.1388J} has a high probability of coming from the $\nu6$ (89\%), whereas L5 Park Forest  \citep{2004M&PS...39.1781B} has a more unclear history with a 48\% 3:1 chance and also significant possible origins from either the $\nu6$ (25\%) or the 5:2 (11\%).
The L chondrite fragments we get today could be the product of smaller and more recent collisions on several present-day parent bodies, and via diverse transport routes.

Outside of the shocked group, \citet{2019M&PS...54..699J} have put forward strong arguments for a source of L chondrites in the inner belt as well.
L5/6 Creston was on an evolved orbit ($a=1.3$\,AU), and has an exceptionally large CRE age of about 40-50 Ma.

In this present work, we report the Desert Fireball Network's latest recovered meteorite fall, an L5 chondrite with an Aten orbit.

This work is layed out with a Data and Methods section, describing the data and data reduction methods used.
Then 4 mostly independent sections follow: trajectory modelling, orbital analyses, darkflight calculations, and circumstances of the recovery.
Finally a conclusions sections highlights the main findings.

\section{Data \& Methods}

\subsection{Astrometric records from photographs}\label{sec:astrom_data}

\begin{figure}
    \centering
    \includegraphics[width=1.\textwidth]{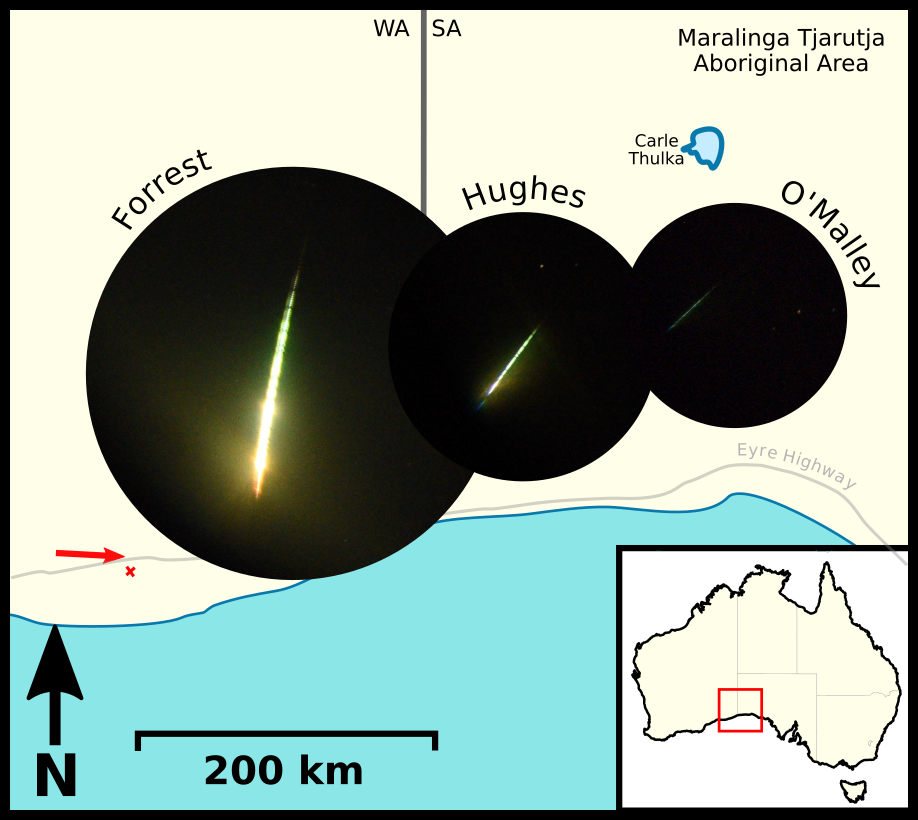}
    \caption{Cropped all-sky images of the fireball from the 3 DFN observatories. Images are of the same pixel scale, 
    with the centre of each image positioned at the observatory location on the map. Dashes encoded in the trajectories are an expression of the liquid crystal shutter modulation and provide both absolute and relative timing along each trajectory. Location of the recovered meteorite near Madura Cave is shown by the red cross.}
    \label{fig:images_on_map}
\end{figure}

On 2020-06-19T20:05:08Z, three Desert Fireball Network camera systems imaged a bright fireball, internally referenced as \textit{\codename} (Table \ref{table:stations}).
The detection was automatically reported to the DFN team by the detection software of \citet{2020PASA...37....8T}.
The location of these are mapped in Figure \ref{fig:images_on_map} along with the fireball as observed by each system.
The 3 camera systems \citep{2017ExA....43..237H} consisted of Nikon D810 digital colour cameras operated at 3200 ISO, and a Samyang 8mm operated at f/4.
A liquid crystal shutter between the lens and the sensor chopped the fireball following a de Bruijn sequence \citep{2017M&PS...52.1669H}.
The resulting point data rate was 10 to 20 samples per second, with each point exposed for 0.01\,s.
Astrometric calibration was performed following the method of \citet{2018M&PS...53.2212D}.
The closest viewpoint, \textit{DFNEXT029} Forrest, had an inferior lens quality, which resulted in astrometric formal uncertainties of $\sim 3$\arcmin\ instead of the nominal $\sim 1-2$\arcmin.
The fireball was also too bright for the shutter breaks to be resolved in the Forrest image between $\sim$38 and $\sim$25\,km altitudes, and at that point too far from other viewpoints.
Dynamical observations from Forrest available below $\sim$25\,km altitude were nonetheless critical to determine how large the main mass was (Section \ref{sec:eks_grits}).

103 data points observed were recorded in total, from just 3 observatories.
The best convergence angle between the 3 observation planes is only 29\degr, while the closest viewpoint to the end of brightflight is located at 170 km, and the other viewpoints are both $>$300 km distant (see Tab. \ref{table:stations}).
These poor observation conditions were partly the result of COVID-19 lockdowns: a number of Nullarbor observatories had not been serviced in 14 months at the time of the fall.

\begin{table}[!h]
	\caption{Locations of Desert Fireball Network Observatories that obtained photographic records of \codename, and nature of data obtained. Times are relative to first fireball observation at \fireballTimeZero\,UTC.
	P: Photographic record (long-exposure high resolution image, see Sec. \ref{sec:astrom_data}), V: compressed PAL video (25 frames per second). Ranges are from when the meteoroid was at 65\,km altitude.
	}              %
	\label{table:stations}    
	\center
	\resizebox{\textwidth}{!}{
	\begin{tabular}{l c c c | c | c c c}          %
		\hline\hline                        %
		 Observatory &&&&Instument& range &  start time & end time \\
		name & latitude & longitude & altitude (m)  & record& (km) & observed & observed \\
		\hline      
		DFNEXT029 - Forrest  &  30.85806 S & 128.11503 E & 166  &P&  200 & 0.00 & 5.50 \\ 
		DFNEXT041 - Hughes & 30.65293 S &  129.70064 E & 144 &P& 336  & 0.92 & 3.12 \\
		DFNSMALL63 - O'Malley &  30.50665 S &  131.19539 E & 122  &P, V&  473  & \tablenotemark{+} &  \tablenotemark{+} \\ 
		\hline                                             %
	\end{tabular}}
		\tablenotetext{+}{shutter breaks were not sufficiently resolved in the still image from O'Malley, this viewpoint was only used for constraining the geometry of the trajectory.}
\end{table}

\subsection{Photometry}\label{sec:photometry}

\begin{figure}[!h]
	\centering
	\includegraphics[width=0.7\textwidth]{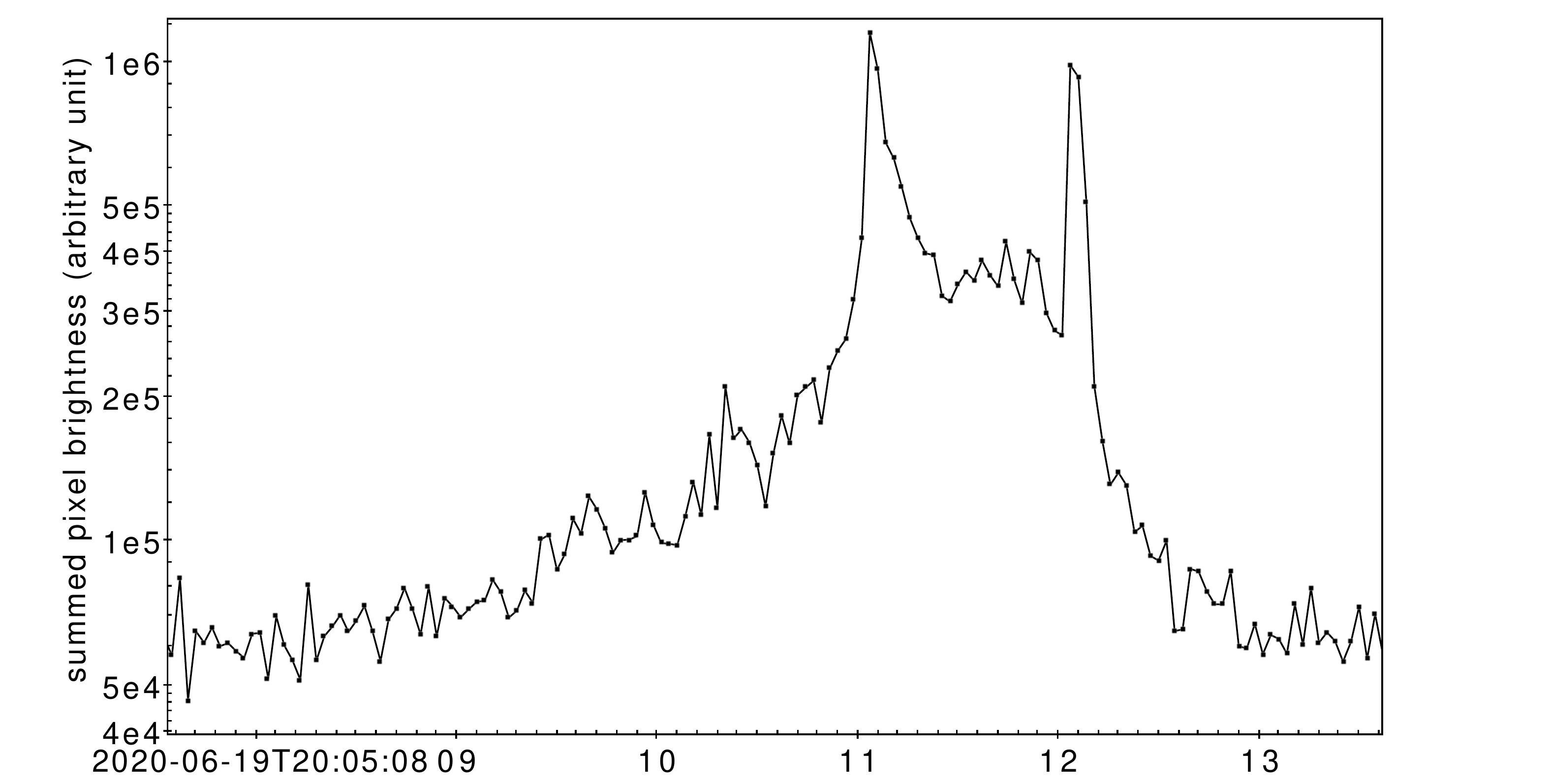}
	\caption{Light curve derived from the video camera of the O'Malley observatory (470 km distant). The two main peaks happened at 2020-06-19T20:05:11.058 and 2020-06-19T20:05:12.058, corresponding to break up heights of 35.8 and 25.8 km, which themselves correspond to ram pressures of $1.5\pm0.1$ and $3.5\pm0.1$ MPa.}
	\label{fig:lc}
\end{figure}

The observatory in O'Malley (Tab. \ref{table:stations}) also recorded a video for the event.
Because of the lossy compression of the video format it is not possible to derive a calibrated light curve from it, but it lets us accurately identify 2 main break up events (Fig. \ref{fig:lc}).
These large fragmentation events are also evident in the still image from Forrest (Fig. \ref{fig:images_on_map}).

The still images from Forrest and Hughes are mostly saturated, and O'Malley is not time resolved (Tab. \ref{table:stations}), therefore we have not used the photographs for photometry purposes.
The results and errors would have been badly constrained, thus having limited value in the analysis.

\section{Trajectory Modelling}\label{sec:modelling_and_orbit}

\subsection{Trajectory determination} \label{sec:triang}

A trajectory of the observed fireball is initially triangulated using the straight line least squares method of \citet{1990BAICz..41..391B}. 
This straight line fit is performed in ECEF coordinates. Although this non inertial reference frame is not ideal, confident timing is not identifiable from the O'Malley image, and we would be unable to include this viewpoint, further decreasing the angle of planes, if triangulation were made in an inertial frame. Conversion of entry velocities and radiants are converted into an inertial frame after the initial triangulation is made.
From this straight line approximation, the fireball trajectory was observed to begin at an altitude of $75.0$ km, with a $58^\circ$ angle to the local horizontal.
The final observation was made at a height of $18.6$ km, having flown a $66.2$ km long trajectory. The two peaks observable in the light curve (Fig. \ref{fig:lc}) at 2020-06-19T20:05:11.058 and 2020-06-19T20:05:12.058 correspond to break up heights of 35.8 and 25.8 km. This indicates breakup at $1.5\pm0.1$ and $3.5\pm0.1$ MPa ram pressures.

Because of the particularly poor observing conditions for this fireball -- notably the low convergence angle -- we create 10,000 Monte Carlo clones of the input data to identify variability in the trajectory solution.
To generate the clones, we randomise the astrometric observations, more or less following the methodology of  \citet{2020MNRAS.491.2688V}, except that the observations are re-sampled in a Gaussian way using the formal astrometric uncertainties, instead of using the residuals to the nominal trajectory fit.
The resulting ensemble of trajectories gives us the inherent variability of the trajectory solution within observation uncertainties. 
For fireballs that have a good convergence angle, and close observing stations, this step is usually not necessary, as Monte Carlo triangulations typically yield solutions within the residuals of the nominal trajectory fit; the triangulation variations are not the main source of uncertainty for meteorite positions nor pre-impact orbit.
In this case, because of the small convergence angle and the distant viewpoints, the Monte Carlo triangulations are required and the ensemble of solutions show a great deal of variability.
The standard errors derived from this ensemble of trajectories attest of the unusually large uncertainty of the trajectory (Table \ref{tab:traj_sum}).
Standard errors on positions are on the order of 200\,m, while the standard error on the direction of the trajectory is $\simeq0.4 \degr$.

\subsection{Estimating initial and terminal masses}\label{sec:eks_grits}

\begin{figure}
    \centering
    \includegraphics[width=0.5\textwidth]{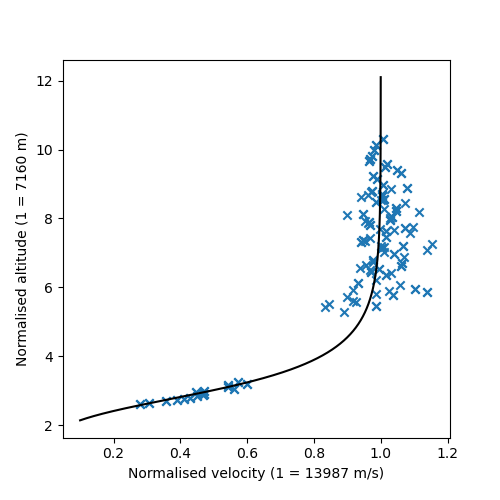}
    \caption{Fitting the normalised velocity-altitude curve to determine the ballistic $\alpha$ -- $\beta$ values.}
    \label{fig:alphabeta}
\end{figure}

We initially use the $\alpha$--$\beta$ criterion to determine if the fireball is a likely meteorite-dropping candidate \citep{2012CosRe..50...56G,2019ApJ...885..115S}. %
The dimensionless ballistic ($\alpha$) and mass loss ($\beta$) parameters calculated for this fireball are $\alpha=9.02$ %
and $\beta=1.03$ %
respectively (Figure \ref{fig:alphabeta}). This positions the event within the likely dropping zone for a 1~kg meteorite (see Github\footnote{https://github.com/desertfireballnetwork/alpha\_beta\_modules}). Although merely a first pass, this method allowed us to quickly establish this was a significant fall and to proceed with further modelling. Assuming meteoroid properties, such as a spherical shape, a bulk density of 3500 \densunitSI, 
and a shape change parameter of 2/3 (see \citealt{2019ApJ...885..115S} and references therein), a terminal mass of 1.3 kg is predicted, with a minimum estimated initial mass of 31 kg. 

We follow this with an Extended Kalman Filter/Smoother, applied to the straight line trajectory \citep{2015M&PS...50.1423S}.
The EK filter is initiated at the end of the observed trajectory ($t_0 + 5.5$\,s), with state values of $3.9\pm0.5$\,\kms for speed and $1\pm1$ kg for mass. 
This method still requires assumed values for meteoroid characteristics, including shape (set to be a rounded brick; $A=1.5$), density ($\rho_m=3500$\,\densunitSI), aerodynamic drag coefficient\footnote{$\Gamma$ is referred to as the drag factor in many meteoroid trajectory works, including \citep{2005M&PS...40...35C} and is related to the aerodynamic drag such that $c_d = 2\Gamma$ \citep{Bronshten1983PhysicsMeteoricPhenomena, 2015aste.book..257B}.} 
($c_d=1$) and apparent ablation coefficient ($\sigma=0.014\,\mbox{s}^{2}\,\mbox{km}^{-2}$; \citealt{2005M&PS...40...35C}).
The filter predicts changes to the state (position, velocity and mass) using the single body aerodynamic equations \citep{2015M&PS...50.1423S}. 
The initial mass and velocity are 64.3 $\pm6$ kg and 13.96$\pm0.07$\kms respectively. Running the subsequent smoother forward in time, we get a final mass and velocity of 2.5$\pm0.6$ kg and 3.76 $\pm0.15$ \kms respectively.

As shown by \citet{2019Icar..321..388S}, the straight line approximation used for the triangulated positions used in these approaches may be an over simplification of the trajectory. Due to the poorly observed fireball however, with few overlapping observations, we are unable to use the 3D particle filter methodology of these authors. We can still perform a particle filter in one dimension applied to the straight line trajectory in this case, to confirm initial velocity and mass estimates having removed meteoroid characteristic assumptions \citep{Sansom2017AnalyzingMeteoroidFlights}. We initialise 10 million particles with values that sample the entire parameter space for these characteristics (see 
\citet{Sansom2017AnalyzingMeteoroidFlights}). Due to the significant fragmentation that occurs between 3-4 seconds (saturating the closest image), the uncertainty in mass at this timestep is increased to 1$\sigma$ = 50\% of the particle mass. This can help estimate the minimum amount of mass lost during this fragmentation event.
The initial mass estimated using this method is $m_0=32 \pm 3$ kg, with a density of 2800 \densunitSI, shape coefficient of A=1.33, $c_d=1$ and apparent ablation coefficient of 0.0101 s~km$^{-2}$. The initial velocity is determined to be $v_0=13.99\pm0.06$\,\kms.

These values are consistent with those calculated using the method of \citet{Gritsevich2007Entrymassbolides, 2009AdSpR..44..323G}%
, despite the characteristic assumptions used in these simpler approaches. It should be noted however that these dynamical methods of estimating initial masses are only able to predict minimum values.

\begin{table}[!h]
	\centering
	\begin{tabular}{llll}
		\hline
		& unit/format & beginning value & end value \\
		\hline
Time & ISO 8601 & 2020-06-19T20:05:07.800 & 2020-06-19T20:05:13.300\\
Latitude & \degr WGS84 & $-31.955 \pm 0.001$  & $-31.974 \pm 0.002$ \\
Longitude & \degr WGS84 & $126.537 \pm 0.002$  & $126.902 \pm 0.002$ \\
Height & m WGS84 & $74981 \pm 174$  & $18628 \pm 101$ \\
Slope & \degr & $58.48 \pm 0.20$  & $58.17 \pm 0.20$ \\
Bearing & \degr & $93.64 \pm 0.58$  & $93.45 \pm 0.58$ \\
Speed  & \ms & $14000 \pm 165$ \textsuperscript{a} & $3756 \pm 146$ \textsuperscript{a}  \\
Apparent radiant (RA) & \degr J2000 & $299.40 \pm 0.31$ \\
Apparent radiant (Dec) & \degr J2000  & $-24.68 \pm 0.21$ \\
    Calculated mass & kg & 
         $64 \pm 6$ \textsuperscript{a}, $31$ \textsuperscript{b}, $29 \pm 13$ \textsuperscript{c}  &
         $2.5 \pm 0.6$ \textsuperscript{a}, $1.3$ \textsuperscript{b} \\
    Calculated density \textsuperscript{c} &\densunitSI&  2800 &\\ 
	\hline
	& unit/format & value & \\
	\hline
		Number of observing stations & & 3 & \\
		Number of data points & & 106 & \\
		Observed duration & s & 5.5 & \\
        Trajectory length & km & 66.2 & \\
        Ballistic parameters \textsuperscript{b} & & $\alpha=9.02$ & $\beta=1.03$ \\
       Apparent ablation coefficient\textsuperscript{c} & s~km$^{-2}$& 0.0101&\\
       Shape coefficient \textsuperscript{c} &&1.33&\\
		Recovery location &\degr WGS84 & lat= -31.96557 & long= 126.98438  \\
		Recovered mass & kg & & 1.072 \\
		\hline
	\end{tabular}
	\caption{Summary of parameters for \codename \meteorite.
	Models used: \textsuperscript{a} \citet{2015M&PS...50.1423S}, \textsuperscript{b} \citet{2009AdSpR..44..323G},
	\textsuperscript{c} \citet{2017AJ....153...87S}.}
	\label{tab:traj_sum}
\end{table}

\section{Orbital modelling}\label{sec:orb}

\subsection{Pre-atmospheric orbit}

\begin{figure}
    \centering
    \includegraphics[width=0.6\textwidth]{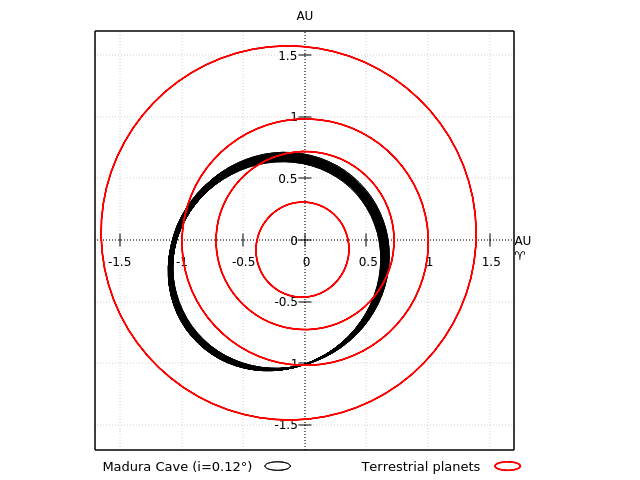}
    \caption{Ecliptic projection representing the pre-encounter orbit of the \meteorite{} meteoroid. Each orbit line corresponds to a Monte Carlo clone from the trajectory calculations.}
    \label{fig:orb}
\end{figure}

Using the integrator of \citet{2019M&PS...54.2149J}, we propagate the position of the meteoroid backwards until it is $10\times$ outside the sphere of influence of the Earth-Moon system.
The positions are then propagated forward to the date of impact, ignoring the influence of the Earth and the Moon.
From this point we convert positions/velocities to ecliptic orbital elements.
Uncertainties are estimated using the 10,000 trajectory clones from Sec. \ref{sec:triang}.
This points to an evolved Aten type orbit for the \meteorite{} meteoroid, with a very low inclination to the ecliptic (Table \ref{tab:orbit} and Figure \ref{fig:orb}).

\begin{table}[!h]
    \centering
    \begin{tabular}{ccc}
    \hline 
    Epoch & TDB & 2020-06-19 \\

Semi-major axis & AU & $0.889 \pm 0.003$ \\
Eccentricity & & $0.327 \pm 0.009$ \\
Inclination & \degr & $0.12 \pm 0.08$ \\
Argument of periapsis & \degr & $312.02\pm 0.51 $ \textsuperscript{*} \\
Longitude ascending node & \degr & 88.70376479 \textsuperscript{*} \\
Perihelion & AU & $0.599 \pm 0.009$ \\
Aphelion & AU & $1.18 \pm 0.007$ \\
Tisserant parameter wrt. Jupiter & & $6.63 \pm 0.02$ \\
Corrected radiant (RA) & \degr & $291.5 \pm 0.4$ \\
Corrected radiant (Dec)  & \degr & $-21.6 \pm 0.3$ \\
Geocentric speed & \ms & $8847 \pm 267$ \\
    \hline 
    \end{tabular}
    \caption{Pre-encounter orbital parameters expressed in the heliocentric ecliptic frame (\textit{J2000}) and associated $1\sigma$ formal uncertainties. \textsuperscript{*}: The uncertainties of argument of perihelion and longitude of ascending node would be large due to low inclination, we therefore fixed the longitude of ascending node to time of impact impact (88.70376479\degr).
}
    \label{tab:orbit}
\end{table}

\subsection{Orbital history}

To better understand the dynamical history of the object in Near-Earth Space, we create 1000 clones of the initial observed vector within the formal uncertainties, and back-track their positions in the past.
This is done following similar methods as \citet{2019AJ....158..183S,2020AJ....159..191S}: the \textit{Rebound} package is used in a simulations where the meteoroid clones evolve under the influence of the 8 planets, the Moon, and the Sun. The simulation is run using the IAS15 adaptive timestep integrator, and the state vector of each particle in the system is recorded every 10,000 years \citep{2015MNRAS.446.1424R}.

We integrated the system backward for 15 million years. At this point, 93\% of the clones are still in the inner Solar System, while 88\% are still in Near-Earth Space ($q<1.3$\,AU). The median orbital elements are telling of this stability: the semi-major shows very slow increase as we move back in time, from 0.9\,AU at the time of impact, to $\simeq$1.1\,AU at 15 Ma, nearly co-orbital with the Earth between 3 and 5 Ma.
We do not integrate further, as 15 million years is already significantly past the Lyapunov timescale in this chaotic part of the Solar System.
No further information can be gained by more prolonged backward integrations.
By fitting an exponential decay function to the number of particles in near-Earth space over time, we find that the NEO dynamical lifetime for such an orbit is $\sim$\,87\,Ma.
We must however stress that these simulations are not to be taken at face value to draw strong conclusions about the dynamical history of Madura Cave.
They merely tell us that Madura Cave has likely spent a long time in NEO space before it impacted the Earth (several tens of million years).

The orbit determined based on the DFN fireball observations (Tab.~\ref{tab:orbit}), is a highly evolved Aten-type.
Amongst meteorite falls, the small semi-major axis  (0.889\,au) is only larger than that of the Bunburra Rockhole meteorite fall \citep{2009Sci...325.1525B}.
The achondritic Bunburra Rockhole was likely transported to near-Earth space via the $\nu_{6}$ resonance.
Given the similarity to Bunburra Rockhole's evolved Aten-type orbit and based on the model described in \citet{2018Icar..312..181G}, \meteorite{} very likely also evolved via the $\nu_{6}$ resonance.

To learn more about the recent thermal environment of \meteorite, we use the simulation results of \citet{2021MNRAS.tmp.1743T} to find out how much time \meteorite{} spent close to the sun.
Their look up table point to a 
52\% probability of the \meteorite{} parent meteoroid having spent some time at perihelion distance $q<0.45$\,AU, for about 0.6 million years in total.
This is more extreme than what most ordinary chondrites would have experienced before their delivery on Earth \citep{2021MNRAS.tmp.1743T}.
Based on the heat model of \citet{2009MNRAS.400..147M}, this could mean that \meteorite{} has been recently heated up to $\sim400$\,K.
The simulations of \citet{2021MNRAS.tmp.1743T} also indicate a non-negligible chance (15\%) of \meteorite{} spending about 0.1 million years at $q<0.3$\,AU, in which case the maximum temperature would have reached $\sim500$\,K.

Using our direct orbital simulations over the last 15 Ma show different results.
We only consider the particles that have not fallen into the Sun (90\% of total).
82\% of the particles have gone below 0.45 AU, and 46\% have been below 0.3\,AU.
The difference with the results from the look up table of \citet{2021MNRAS.tmp.1743T} is not surprising, as the ($a,e,i$) orbital elements alone are not fully descriptive of the history of a particular orbit.
So \meteorite{} is actually quite likely to have been heated up to over 500\,K.

\section{Darkflight and wind modelling}

\begin{figure}
    \centering
    \includegraphics[width=0.6\textwidth]{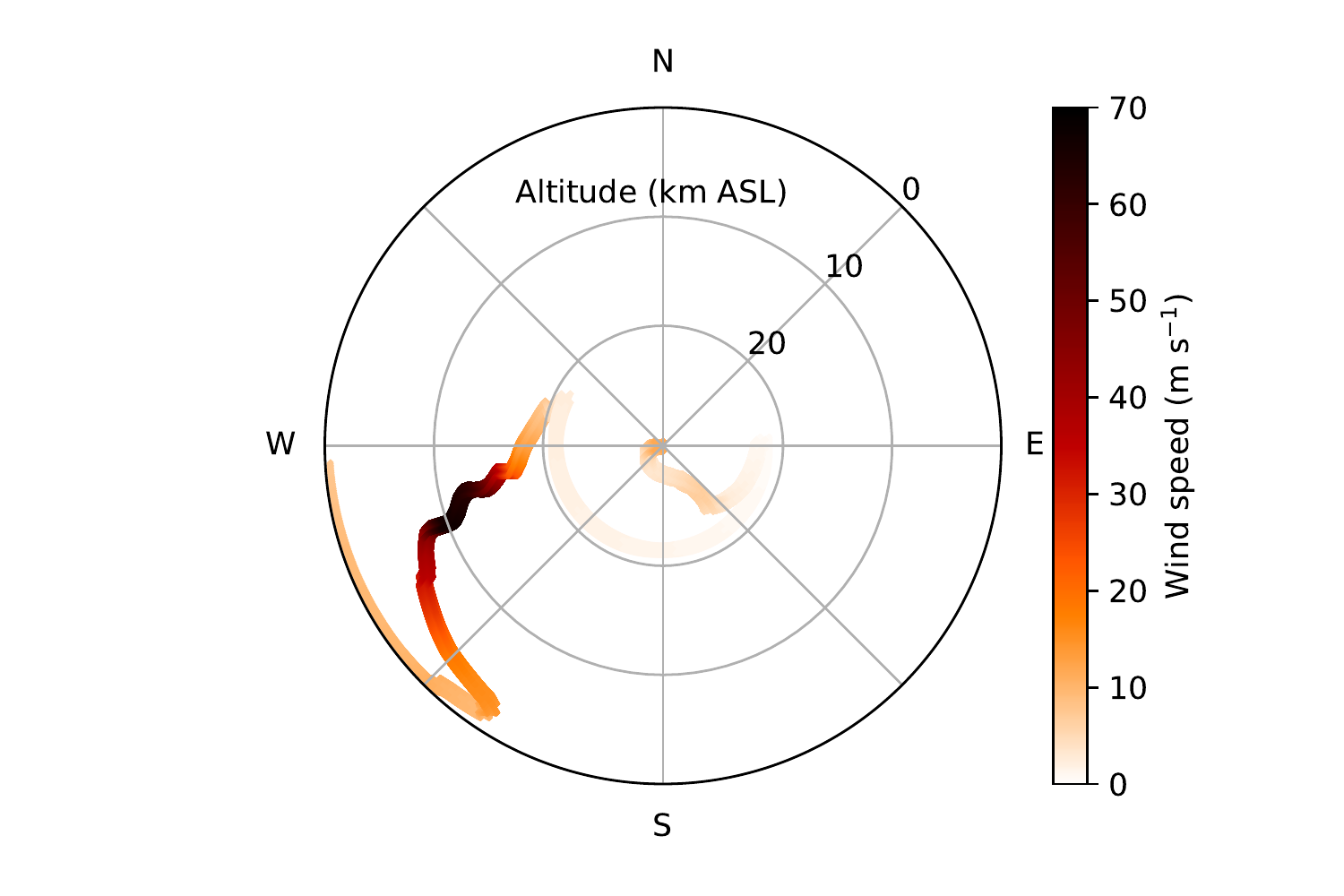}
    \caption{Wind model (speed and direction for a given altitude), extracted as a vertical profile at the coordinates of the lowest visible bright flight measurement at 2020-06-19T20:05:08Z. Model integration started at 2020-06-19T12:00. A strong jet stream was present at the time of the fall, with a maximum of  $\sim$67\,\ms around 10.6 km altitude, coming from a W-SW direction. Data file is available as supplementary material.}
    \label{fig:Wind}
\end{figure}

We modelled the atmospheric conditions numerically using the Weather Research and Forecasting (WRF) model version 4.0 with dynamic solver ARW (Advanced Research WRF)  \citep{Skamarock2019descriptionadvancedresearch}.
The weather model (Figure \ref{fig:Wind}) includes wind speed, wind direction, pressure, temperature and relative humidity at heights ranging up to 30 km.
We did 4 runs, starting the weather simulation at different times before the meteorite fall (on 2020-06-19 at 0:00, 6:00, 12:00 and 18:00).
In this instance all 4 models give relatively similar profiles, which signals a stable weather situation (this is not always the case \citep{2018M&PS...53.2212D}).
Fig. \ref{fig:Wind} shows a 1D vertical section of one of the models, defined by the location of the calculated end point of the bright flight.
The data tables of all the models are available as supplementary material.

\begin{figure}
    \centering
    \includegraphics[width=1.\textwidth]{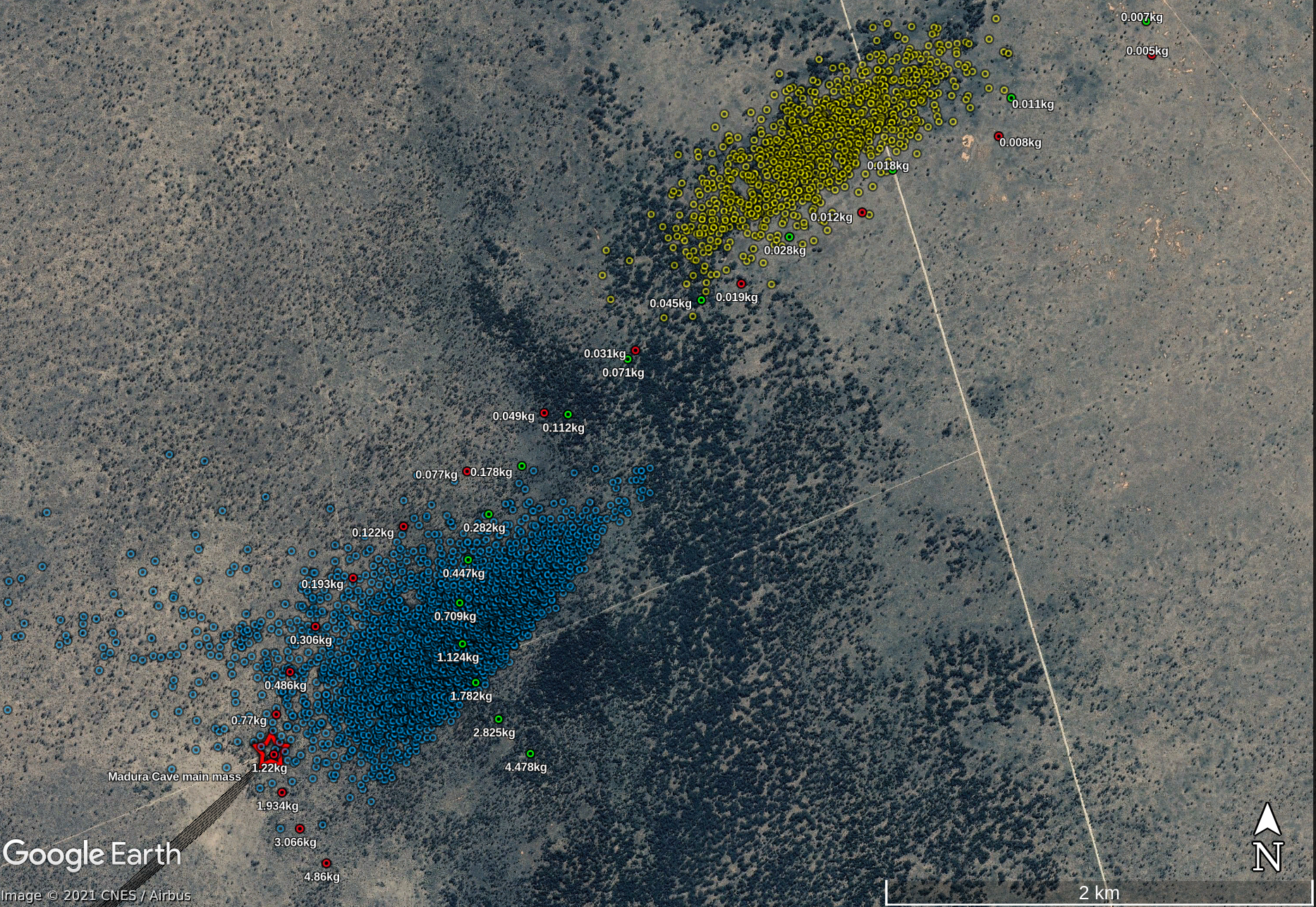}
    \caption{Fall area of Madura Cave.
    Red star marks the recovered main mass location.
    Blue dots: Ground positions of a 1.1 kg mass from the 10,000 Monte Carlo simulations from Sec. \ref{sec:triang}.
    Red dots: fall line that corresponds to the blue dot best matching the meteorite found.
    Green dots: fall line for the nominal trajectory.
    Yellow dots: Monte Carlo simulations for the possible location of fragments from the second fragmentation event visible in the light curve.}
    \label{fig:fall_line}
\end{figure}

Using these atmosphere models, we can then propagate bright flight observations to the ground using the dark flight model of \citet{2021arXiv210804397T}.
In Figure \ref{fig:fall_line}, we illustrate the various factors at play that drive the uncertainty of the fall locations.
We propagated a 1.2\,kg mass (assuming a cylindrical shape) to the ground from each of the 10,000 Monte Carlo simulations from Sec. \ref{sec:triang}.
This 1.2\,kg mass roughly corresponds to what the main mass must have been at the last observation point we had.
The resulting impact points (blue dots in Fig. \ref{fig:fall_line}) illustrate at which point the variability of the triangulation affects the fall locations.
The main mass was found within the cloud of points, but somewhat far from the fall location taken from the nominal fall line (green dots).
This indicates that our error analysis is adequate, but this would not have been a comfortable situation if we had had to search the entire area in order to find the meteorite.
Had we not been fortunate in quickly locating the stone (Sec. \ref{sec:search}), the area to search would have been around 5\,km$^2$, instead of $\sim$0.5\,km$^2$ if the fireball had been well observed (in which case the ground error would have been dominated by the uncertainty in mass and shape).

As the fireball showed 2 significant late peaks in its light curve (Sec. \ref{sec:photometry}), it is reasonable to assume that debris would have emanated from these fragmentation events.
The first of these happened at 36\,km altitude, significantly outside of our weather model coverage (max $30.7$\,km).
On the other hand, the second peak happened much lower around 26\,km altitude, so we can predict where the resulting debris could have landed.
Using the best meteorite matching trajectory (red fall line in Fig. \ref{fig:fall_line}), we perform a Monte Carlo simulations of 1000 particles from the second fragmentation point, varying the parameters as such: 
10 grams $\pm$ 50\%, 3500$\pm$500 \densunitSI, drag of a sphere $\pm$ 10\%, and wind magnitude uncertainties of $\pm$2.0 \ms.
Fortunately, the choice of the trajectory for this small masses simulation has much less drastic effect on the ground locations, as in this case the fall lines converge at the low mass end.
This gives us a cloud of points (yellow dots in Fig. \ref{fig:fall_line}) that represents where it should be possible to find fragments.

\section{Search and recovery}\label{sec:search}

\begin{figure}
  \centering

  \vspace{\floatsep}

  \begin{tabular}{@{}c@{}}
    \includegraphics[width=0.4\textwidth]{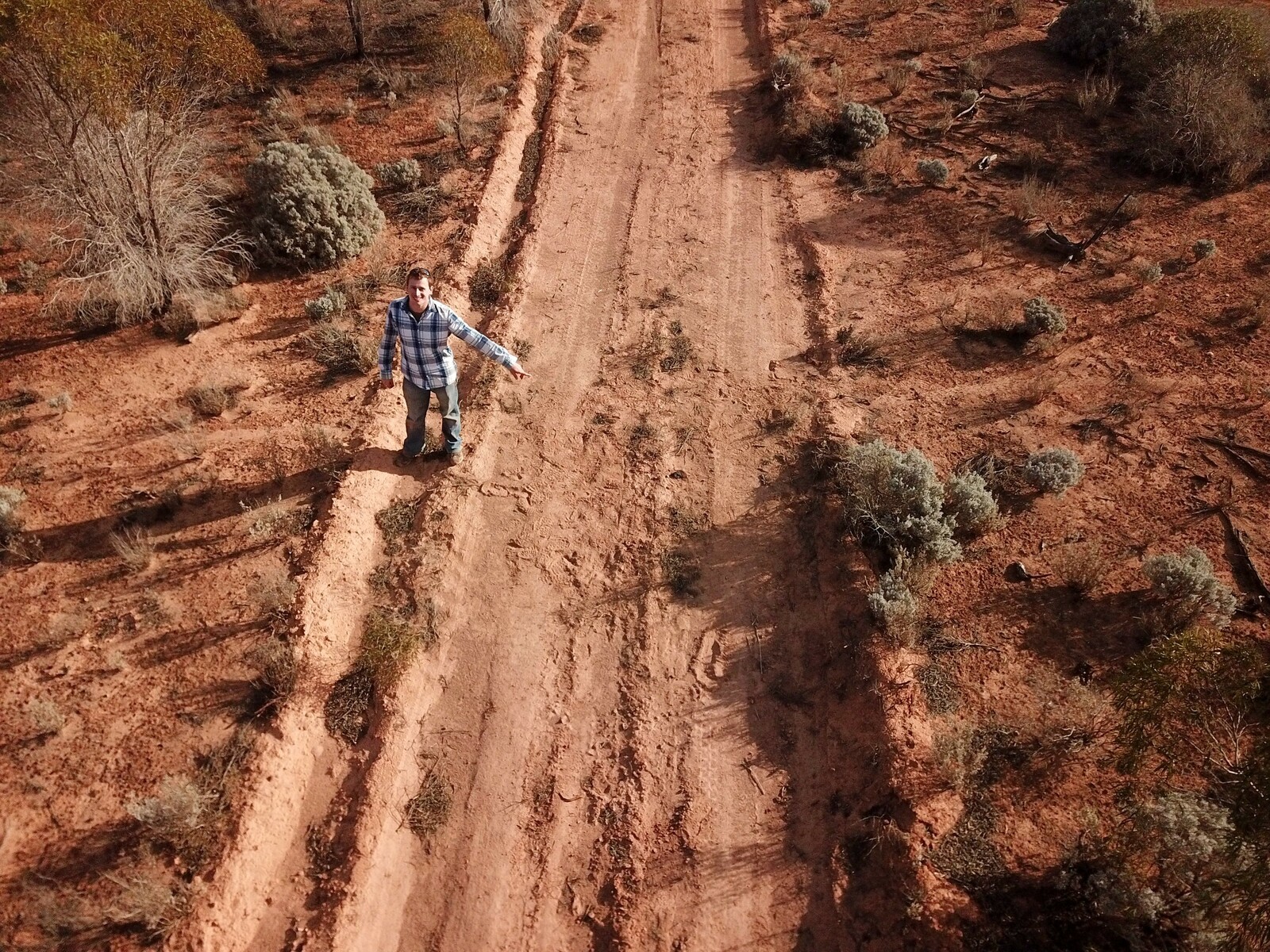}
    \small (a) Old telegraph track on which Madura Cave was found.
  \end{tabular}
  
   \vspace{\floatsep}

  \begin{tabular}{@{}c@{}}
    \includegraphics[width=0.4\textwidth]{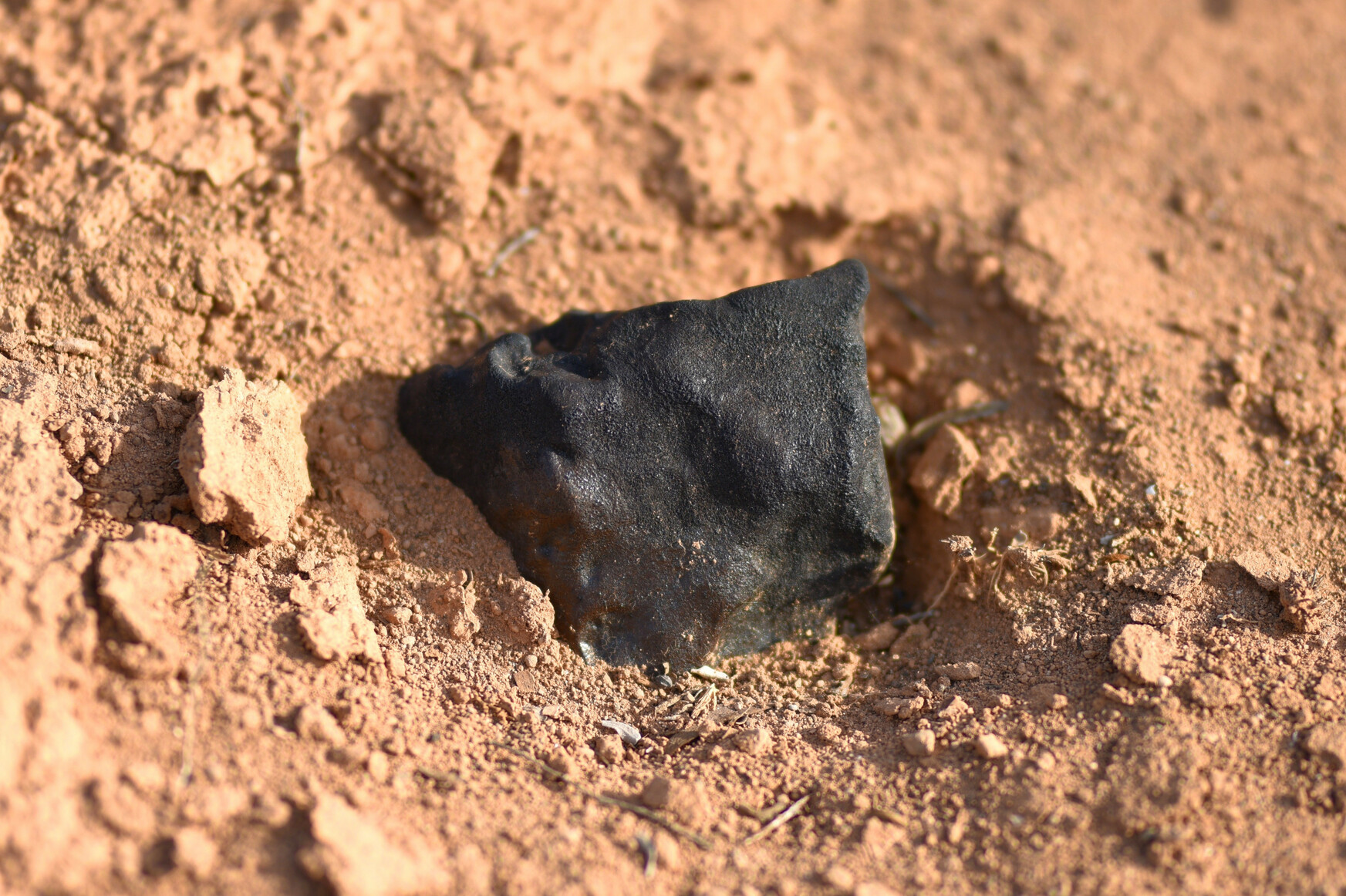}
    \small (b) Madura Cave (1.072 kg $\sim 11 \times 9 \times 8$\,cm).
  \end{tabular}
   \vspace{\floatsep}

  \begin{tabular}{@{}c@{}}
    \includegraphics[width=0.4\textwidth]{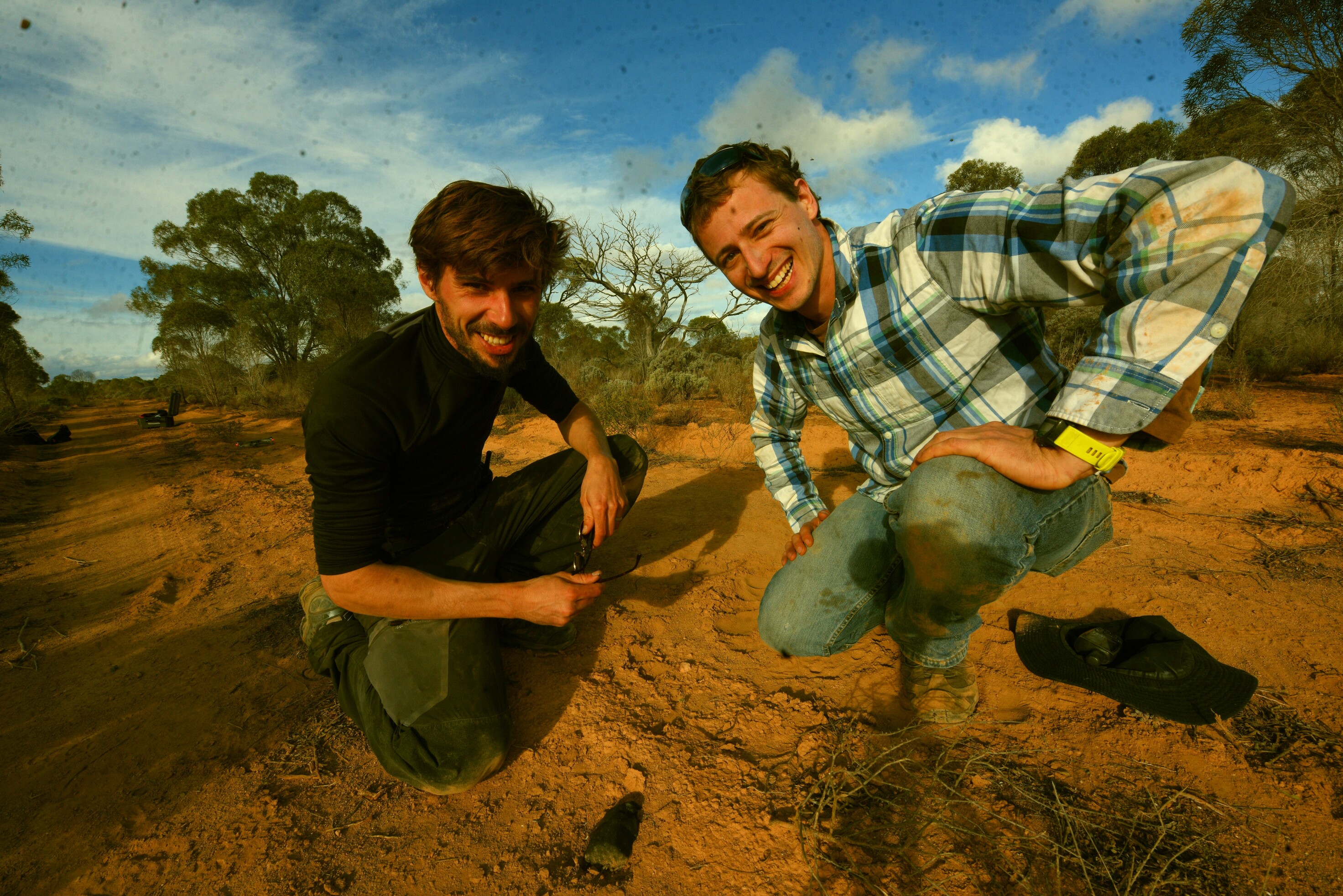}
    \small (c) AL and HD in front of their find.
  \end{tabular}
  
  \vspace{\floatsep}

  \begin{tabular}{@{}c@{}}
    \includegraphics[width=0.4\textwidth]{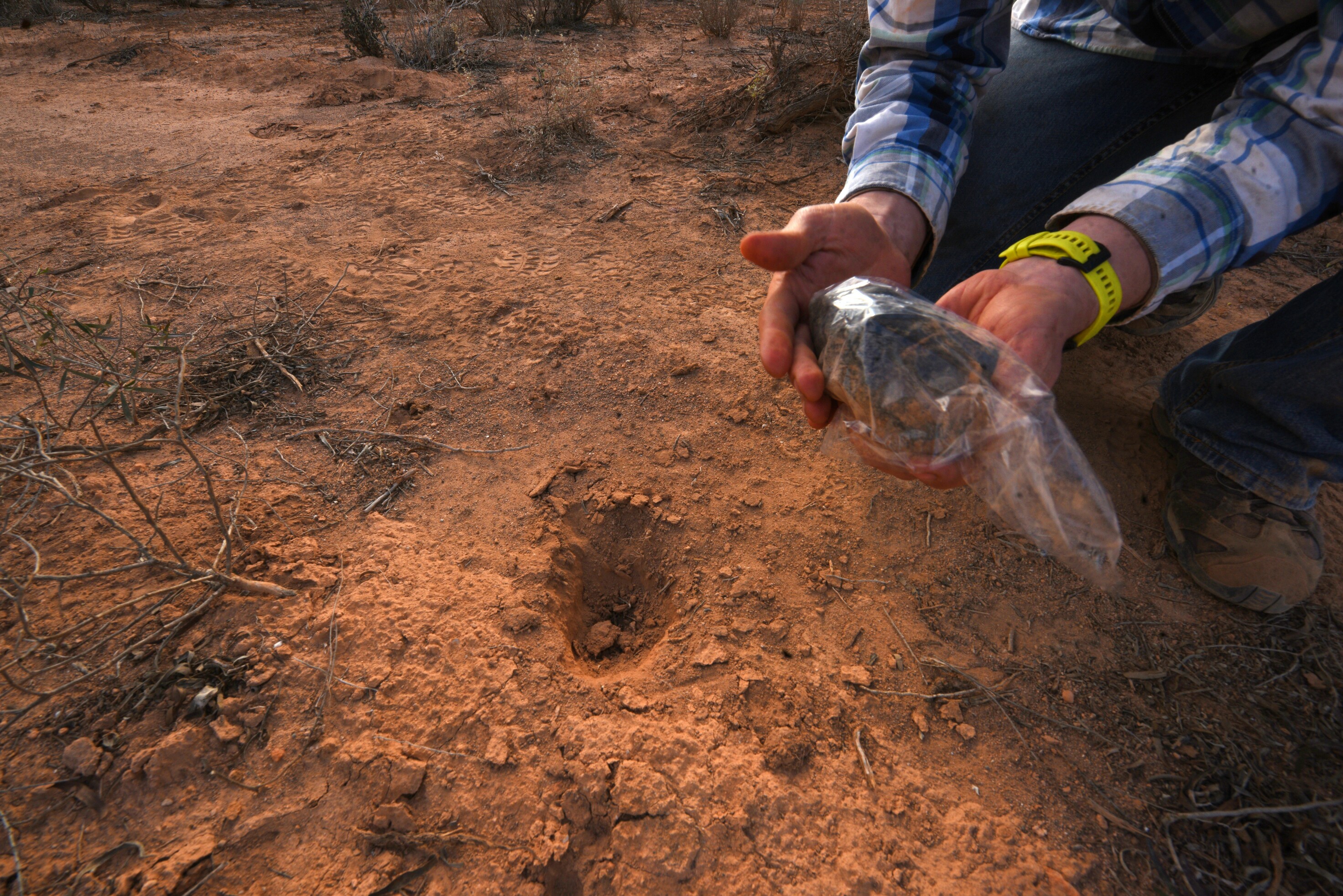}
    \small (d) Collection of the meteorite in a Teflon bag.
  \end{tabular}

  \caption{Recovery of \meteorite{} main mass.}
  \label{fig:search}
\end{figure}

In early June 2020, once the strictest Covid restrictions were lifted and travel was allowed, the DFN team was getting ready to send a team to search what would eventually become the Mundrabilla Fault meteorite.
However when the present fireball happened on June 20th, priority was given to it as it was a larger main mass, hence easier to find.
Observational data was scarce however (Sec. \ref{sec:triang}), as some of the closest observatories had gone offline.
Not knowing if they were offline because of an internet connection fault or a more serious matter, HD and AL planned a short trip to visit some of these observatories (Kanandah, Kybo, and Mundrabilla), with the hope of refining the fall area predictions with the extra viewpoints.
They discovered that each camera station suffered major faults and therefore did not capture data.
The team nonetheless spent one day at the fall site on their way back to Perth, on 2020-07-09.
This detour was meant to collect drone training images for automated meteorite searching \citep{2020M&PS...55.2461A}.
HD and AL also walked the predicted fall area of the main mass, in order to assess the quality of the searching ground for their colleagues.
When walking back to their vehicle along a track, they stumbled upon the main mass (Fig. \ref{fig:search}), just 19 days after the fall.
The \meteorite{} main mass (1.072\,kg) was found at coordinates  \href{geo:-31.96557,126.98438}{$\phi = -31.96557 $ $\lambda = 126.98438$}.
It is believed that at least some rain has fallen on the rock before recovery: the nearby weather observation station in Eucla ($\sim$200 km away) recorded 10 rainy days out of the 19 days period (source: Bureau of Meteorology).
Most of these rain episodes were light though, with 1.2\,mm being the daily maximum recorded on 2020-06-22, so the area is unlikely to have been flooded while the meteorite was on the ground.

As of August 2021, the fall area of the fragments has not been searched (Fig. \ref{fig:fall_line}).

\section{Conclusions}

The orbit \meteorite{} was on before impact suggests the meteoroid is likely to have spent tens of millions of years in near-Earth space.

The low-inclination Aten orbit is also characteristic of an inner main belt origin, via the $\nu6$ resonance.
This would confirm the presence of an L chondrite present-day parent body in the inner main belt, as first suggested by \citet{2019M&PS...54..699J}.
Whether \meteorite{} and \textit{Creston} are connected, from the same present-day parent body, or even maybe from the same ejecting impact, will have to be investigated via rock dating analyses.

This will be the subject of a future study, but based on Madura Cave's large NEO dynamical lifetime we should expect its cosmic ray exposure age to be relatively old for an ordinary chondrite.

\section{Supplementary material}

Supplementary material has been uploaded as a \textit{Zenodo} record at \url{http://doi.org/10.5281/zenodo.5763497}.
It contains fireball images (including calibration data), astrometry tables, the trajectory, wind profiles, as well as images of the recovery.

\begin{acknowledgements}

This work was funded by the Australian Research Council as part of the Australian Discovery Project scheme (DP170102529, DP200102073), and receives institutional support from Curtin University. Data reduction is supported by resources provided by the Pawsey Supercomputing Centre with funding from the Australian Government and the Government of Western Australia.
The DFN data reduction pipeline makes intensive use of Astropy, a community-developed core Python package for Astronomy \citep{2013A&A...558A..33A}.
\end{acknowledgements}

\bibliography{biblio}

\begin{thebibliography}{}
\expandafter\ifx\csname natexlab\endcsname\relax\def\natexlab#1{#1}\fi
\providecommand{\url}[1]{\href{#1}{#1}}
\providecommand{\dodoi}[1]{doi:~\href{http://doi.org/#1}{\nolinkurl{#1}}}
\providecommand{\doeprint}[1]{\href{http://ascl.net/#1}{\nolinkurl{http://ascl.net/#1}}}
\providecommand{\doarXiv}[1]{\href{https://arxiv.org/abs/#1}{\nolinkurl{https://arxiv.org/abs/#1}}}

\bibitem[{{Anderson} {et~al.}(2020){Anderson}, {Towner}, {Bland}, {Haikings},
  {Volante}, {Sansom}, {Devillepoix}, {Shober}, {Hartig}, {Cupak},
  {Jansen-Sturgeon}, {Howie}, {Benedix}, \& {Deacon}}]{2020M&PS...55.2461A}
{Anderson}, S., {Towner}, M., {Bland}, P., {et~al.} 2020, \maps, 55, 2461,
  \dodoi{10.1111/maps.13593}

\bibitem[{{Astropy Collaboration} {et~al.}(2013){Astropy Collaboration},
  {Robitaille}, {Tollerud}, {Greenfield}, {Droettboom}, {Bray}, {Aldcroft},
  {Davis}, {Ginsburg}, {Price-Whelan}, {Kerzendorf}, {Conley}, {Crighton},
  {Barbary}, {Muna}, {Ferguson}, {Grollier}, {Parikh}, {Nair}, {Unther},
  {Deil}, {Woillez}, {Conseil}, {Kramer}, {Turner}, {Singer}, {Fox}, {Weaver},
  {Zabalza}, {Edwards}, {Azalee Bostroem}, {Burke}, {Casey}, {Crawford},
  {Dencheva}, {Ely}, {Jenness}, {Labrie}, {Lim}, {Pierfederici}, {Pontzen},
  {Ptak}, {Refsdal}, {Servillat}, \& {Streicher}}]{2013A&A...558A..33A}
{Astropy Collaboration}, {Robitaille}, T.~P., {Tollerud}, E.~J., {et~al.} 2013,
  \aap, 558, A33, \dodoi{10.1051/0004-6361/201322068}

\bibitem[{{Bland} {et~al.}(2009){Bland}, {Spurn{\'y}}, {Towner}, {Bevan},
  {Singleton}, {Bottke}, {Greenwood}, {Chesley}, {Shrben{\'y}},
  {Borovi{\v{c}}ka}, {Ceplecha}, {McClafferty}, {Vaughan}, {Benedix}, {Deacon},
  {Howard}, {Franchi}, \& {Hough}}]{2009Sci...325.1525B}
{Bland}, P.~A., {Spurn{\'y}}, P., {Towner}, M.~C., {et~al.} 2009, Science, 325,
  1525, \dodoi{10.1126/science.1174787}

\bibitem[{{Borovi{\v c}ka}(1990)}]{1990BAICz..41..391B}
{Borovi{\v c}ka}, J. 1990, Bulletin of the Astronomical Institutes of
  Czechoslovakia, 41, 391

\bibitem[{{Borovi{\v c}ka} {et~al.}(2015){Borovi{\v c}ka}, {Spurn{\'y}}, \&
  {Brown}}]{2015aste.book..257B}
{Borovi{\v c}ka}, J., {Spurn{\'y}}, P., \& {Brown}, P. 2015, {Small Near-Earth
  Asteroids as a Source of Meteorites}, ed. P.~{Michel}, F.~E. {DeMeo}, \&
  W.~F. {Bottke} (University of Arizona Press), 257--280,
  \dodoi{10.2458/azu_uapress_9780816532131-ch014}

\bibitem[{Bronshten(1983)}]{Bronshten1983PhysicsMeteoricPhenomena}
Bronshten, V.~A. 1983, {Physics of Meteoric Phenomena}, Geophysics and
  Astrophysics Monographs (Dordrecht, Netherlands: Reidel)

\bibitem[{{Brown} {et~al.}(2004){Brown}, {Pack}, {Edwards}, {Revelle}, {Yoo},
  {Spalding}, \& {Tagliaferri}}]{2004M&PS...39.1781B}
{Brown}, P., {Pack}, D., {Edwards}, W.~N., {et~al.} 2004, Meteoritics and
  Planetary Science, 39, 1781, \dodoi{10.1111/j.1945-5100.2004.tb00075.x}

\bibitem[{{Ceplecha} \& {Revelle}(2005)}]{2005M&PS...40...35C}
{Ceplecha}, Z., \& {Revelle}, D.~O. 2005, \maps, 40, 35,
  \dodoi{10.1111/j.1945-5100.2005.tb00363.x}

\bibitem[{{Devillepoix} {et~al.}(2018){Devillepoix}, {Sansom}, {Bland},
  {Towner}, {Cup{\'a}K}, {Howie}, {Jansen-Sturgeon}, {Cox}, {Hartig},
  {Benedix}, \& {Paxman}}]{2018M&PS...53.2212D}
{Devillepoix}, H. A.~R., {Sansom}, E.~K., {Bland}, P.~A., {et~al.} 2018, \maps,
  53, 2212, \dodoi{10.1111/maps.13142}

\bibitem[{{Granvik} \& {Brown}(2018)}]{2018Icar..311..271G}
{Granvik}, M., \& {Brown}, P. 2018, \icarus, 311, 271,
  \dodoi{10.1016/j.icarus.2018.04.012}

\bibitem[{{Granvik} {et~al.}(2018){Granvik}, {Morbidelli}, {Jedicke}, {Bolin},
  {Bottke}, {Beshore}, {Vokrouhlick{\'y}}, {Nesvorn{\'y}}, \&
  {Michel}}]{2018Icar..312..181G}
{Granvik}, M., {Morbidelli}, A., {Jedicke}, R., {et~al.} 2018, \icarus, 312,
  181, \dodoi{10.1016/j.icarus.2018.04.018}

\bibitem[{{Gritsevich}(2009)}]{2009AdSpR..44..323G}
{Gritsevich}, M.~I. 2009, Advances in Space Research, 44, 323,
  \dodoi{10.1016/j.asr.2009.03.030}

\bibitem[{Gritsevich \& Stulov(2007)}]{Gritsevich2007Entrymassbolides}
Gritsevich, M.~I., \& Stulov, V.~P. 2007, Doklady Physics, 52, 219,
  \dodoi{10.1134/S102833580704012X}

\bibitem[{{Gritsevich} {et~al.}(2012){Gritsevich}, {Stulov}, \&
  {Turchak}}]{2012CosRe..50...56G}
{Gritsevich}, M.~I., {Stulov}, V.~P., \& {Turchak}, L.~I. 2012, Cosmic
  Research, 50, 56, \dodoi{10.1134/S0010952512010017}

\bibitem[{{Howie} {et~al.}(2017{\natexlab{a}}){Howie}, {Paxman}, {Bland},
  {Towner}, {Cupak}, {Sansom}, \& {Devillepoix}}]{2017ExA....43..237H}
{Howie}, R.~M., {Paxman}, J., {Bland}, P.~A., {et~al.} 2017{\natexlab{a}},
  Experimental Astronomy, 43, 237, \dodoi{10.1007/s10686-017-9532-7}

\bibitem[{{Howie} {et~al.}(2017{\natexlab{b}}){Howie}, {Paxman}, {Bland},
  {Towner}, {Sansom}, \& {Devillepoix}}]{2017M&PS...52.1669H}
---. 2017{\natexlab{b}}, \maps, 52, 1669, \dodoi{10.1111/maps.12878}

\bibitem[{{Jansen-Sturgeon} {et~al.}(2019){Jansen-Sturgeon}, {Sansom}, \&
  {Bland}}]{2019M&PS...54.2149J}
{Jansen-Sturgeon}, T., {Sansom}, E.~K., \& {Bland}, P.~A. 2019, Meteoritics and
  Planetary Science, 54, 2149, \dodoi{10.1111/maps.13376}

\bibitem[{{Jenniskens}(2020)}]{2020IAUGA..30....9J}
{Jenniskens}, P. 2020, in IAU General Assembly, 9--12,
  \dodoi{10.1017/S1743921319003235}

\bibitem[{{Jenniskens} {et~al.}(2014){Jenniskens}, {Rubin}, {Yin}, {Sears},
  {Sandford}, {Zolensky}, {Krot}, {Blair}, {Kane}, {Utas}, {Verish},
  {Friedrich}, {Wimpenny}, {Eppich}, {Ziegler}, {Verosub}, {Rowland}, {Albers},
  {Gural}, {Grigsby}, {Fries}, {Matson}, {Johnston}, {Silber}, {Brown},
  {Yamakawa}, {Sanborn}, {Laubenstein}, {Welten}, {Nishiizumi}, {Meier},
  {Busemann}, {Clay}, {Caffee}, {Schmitt-Kopplin}, {Hertkorn}, {Glavin},
  {Callahan}, {Dworkin}, {Wu}, {Zare}, {Grady}, {Verchovsky}, {Emel'Yanenko},
  {Naroenkov}, {Clark}, {Girten}, \& {Worden}}]{2014M&PS...49.1388J}
{Jenniskens}, P., {Rubin}, A.~E., {Yin}, Q.-Z., {et~al.} 2014, Meteoritics and
  Planetary Science, 49, 1388, \dodoi{10.1111/maps.12323}

\bibitem[{{Jenniskens} {et~al.}(2019){Jenniskens}, {Utas}, {Yin}, {Matson},
  {Fries}, {Howell}, {Free}, {Albers}, {Devillepoix}, {Bland}, {Miller},
  {Verish}, {Garvie}, {Zolensky}, {Ziegler}, {Sanborn}, {Verosub}, {Rowland},
  {Ostrowski}, {Bryson}, {Laubenstein}, {Zhou}, {Li}, {Li}, {Liu}, {Tang},
  {Welten}, {Caffee}, {Meier}, {Plant}, {Maden}, {Busemann}, {Granvik}, \&
  {Creston Meteorite Consortium}}]{2019M&PS...54..699J}
{Jenniskens}, P., {Utas}, J., {Yin}, Q.-Z., {et~al.} 2019, \maps, 54, 699,
  \dodoi{10.1111/maps.13235}

\bibitem[{{Marchi} {et~al.}(2009){Marchi}, {Delbo'}, {Morbidelli}, {Paolicchi},
  \& {Lazzarin}}]{2009MNRAS.400..147M}
{Marchi}, S., {Delbo'}, M., {Morbidelli}, A., {Paolicchi}, P., \& {Lazzarin},
  M. 2009, \mnras, 400, 147, \dodoi{10.1111/j.1365-2966.2009.15459.x}

\bibitem[{{McGraw} {et~al.}(2018){McGraw}, {Reddy}, \&
  {Sanchez}}]{2018MNRAS.476..630M}
{McGraw}, A.~M., {Reddy}, V., \& {Sanchez}, J.~A. 2018, \mnras, 476, 630,
  \dodoi{10.1093/mnras/sty250}

\bibitem[{{Nesvorn{\'y}} {et~al.}(2009){Nesvorn{\'y}}, {Vokrouhlick{\'y}},
  {Morbidelli}, \& {Bottke}}]{2009Icar..200..698N}
{Nesvorn{\'y}}, D., {Vokrouhlick{\'y}}, D., {Morbidelli}, A., \& {Bottke},
  W.~F. 2009, \icarus, 200, 698, \dodoi{10.1016/j.icarus.2008.12.016}

\bibitem[{{Rein} \& {Spiegel}(2015)}]{2015MNRAS.446.1424R}
{Rein}, H., \& {Spiegel}, D.~S. 2015, \mnras, 446, 1424,
  \dodoi{10.1093/mnras/stu2164}

\bibitem[{{Sansom} {et~al.}(2015){Sansom}, {Bland}, {Paxman}, \&
  {Towner}}]{2015M&PS...50.1423S}
{Sansom}, E.~K., {Bland}, P., {Paxman}, J., \& {Towner}, M. 2015, Meteoritics
  and Planetary Science, 50, 1423, \dodoi{10.1111/maps.12478}

\bibitem[{{Sansom} {et~al.}(2017{\natexlab{a}}){Sansom}, {Rutten}, \&
  {Bland}}]{Sansom2017AnalyzingMeteoroidFlights}
{Sansom}, E.~K., {Rutten}, M.~G., \& {Bland}, P.~A. 2017{\natexlab{a}}, \aj,
  153, 87, \dodoi{10.3847/1538-3881/153/2/87}

\bibitem[{{Sansom} {et~al.}(2017{\natexlab{b}}){Sansom}, {Rutten}, \&
  {Bland}}]{2017AJ....153...87S}
---. 2017{\natexlab{b}}, \aj, 153, 87, \dodoi{10.3847/1538-3881/153/2/87}

\bibitem[{{Sansom} {et~al.}(2019{\natexlab{a}}){Sansom}, {Gritsevich},
  {Devillepoix}, {Jansen-Sturgeon}, {Shober}, {Bland}, {Towner}, {Cup{\'a}k},
  {Howie}, \& {Hartig}}]{2019ApJ...885..115S}
{Sansom}, E.~K., {Gritsevich}, M., {Devillepoix}, H. A.~R., {et~al.}
  2019{\natexlab{a}}, \apj, 885, 115, \dodoi{10.3847/1538-4357/ab4516}

\bibitem[{{Sansom} {et~al.}(2019{\natexlab{b}}){Sansom}, {Jansen-Sturgeon},
  {Rutten}, {Devillepoix}, {Bland}, {Howie}, {Cox}, {Towner}, {Cup{\'a}k}, \&
  {Hartig}}]{2019Icar..321..388S}
{Sansom}, E.~K., {Jansen-Sturgeon}, T., {Rutten}, M.~G., {et~al.}
  2019{\natexlab{b}}, \icarus, 321, 388, \dodoi{10.1016/j.icarus.2018.09.026}

\bibitem[{{Schmitz} {et~al.}(2001){Schmitz}, {Tassinari}, \&
  {Peucker-Ehrenbrink}}]{2001E&PSL.194....1S}
{Schmitz}, B., {Tassinari}, M., \& {Peucker-Ehrenbrink}, B. 2001, Earth and
  Planetary Science Letters, 194, 1, \dodoi{10.1016/S0012-821X(01)00559-3}

\bibitem[{{Shober} {et~al.}(2019){Shober}, {Jansen-Sturgeon}, {Sansom},
  {Devillepoix}, {Bland}, {Cup{\'a}k}, {Towner}, {Howie}, \&
  {Hartig}}]{2019AJ....158..183S}
{Shober}, P.~M., {Jansen-Sturgeon}, T., {Sansom}, E.~K., {et~al.} 2019, \aj,
  158, 183, \dodoi{10.3847/1538-3881/ab3f2d}

\bibitem[{{Shober} {et~al.}(2020){Shober}, {Jansen-Sturgeon}, {Sansom},
  {Devillepoix}, {Towner}, {Bland}, {Cup{\'a}k}, {Howie}, \&
  {Hartig}}]{2020AJ....159..191S}
---. 2020, \aj, 159, 191, \dodoi{10.3847/1538-3881/ab8002}

\bibitem[{{Skamarock} {et~al.}(2019){Skamarock}, {Klemp}, {Dudhia}, {Gill},
  {Liu}, {Berner}, {Wang}, {Powers}, {Duda}, {Barker}, \&
  {Huang}}]{Skamarock2019descriptionadvancedresearch}
{Skamarock}, W.~C., {Klemp}, J.~B., {Dudhia}, J., {et~al.} 2019, A description
  of the advanced research WRF version 4, Tech. rep., NCAR Technical Note
  NCAR/TN-556+STR, \dodoi{doi:10.5065/1dfh-6p97}

\bibitem[{{Spoto} {et~al.}(2015){Spoto}, {Milani}, \&
  {Kne{\v{z}}evi{\'c}}}]{2015Icar..257..275S}
{Spoto}, F., {Milani}, A., \& {Kne{\v{z}}evi{\'c}}, Z. 2015, \icarus, 257, 275,
  \dodoi{10.1016/j.icarus.2015.04.041}

\bibitem[{{Toliou} {et~al.}(2021){Toliou}, {Granvik}, \&
  {Tsirvoulis}}]{2021MNRAS.tmp.1743T}
{Toliou}, A., {Granvik}, M., \& {Tsirvoulis}, G. 2021, \mnras,
  \dodoi{10.1093/mnras/stab1934}

\bibitem[{{Towner} {et~al.}(2020){Towner}, {Cupak}, {Deshayes}, {Howie},
  {Hartig}, {Paxman}, {Sansom}, {Devillepoix}, {Jansen-Sturgeon}, \&
  {Bland}}]{2020PASA...37....8T}
{Towner}, M.~C., {Cupak}, M., {Deshayes}, J., {et~al.} 2020, \pasa, 37, e008,
  \dodoi{10.1017/pasa.2019.48}

\bibitem[{{Towner} {et~al.}(2021){Towner}, {Jansen-Sturgeon}, {Cupak},
  {Sansom}, {Devillepoix}, {Bland}, {Howie}, {Paxman}, {Benedix}, \&
  {Hartig}}]{2021arXiv210804397T}
{Towner}, M.~C., {Jansen-Sturgeon}, T., {Cupak}, M., {et~al.} 2021, arXiv
  e-prints, arXiv:2108.04397.
\newblock \doarXiv{2108.04397}

\bibitem[{{Vernazza} {et~al.}(2008){Vernazza}, {Binzel}, {Thomas}, {DeMeo},
  {Bus}, {Rivkin}, \& {Tokunaga}}]{2008Natur.454..858V}
{Vernazza}, P., {Binzel}, R.~P., {Thomas}, C.~A., {et~al.} 2008, \nat, 454,
  858, \dodoi{10.1038/nature07154}

\bibitem[{{Vida} {et~al.}(2020){Vida}, {Gural}, {Brown}, {Campbell-Brown}, \&
  {Wiegert}}]{2020MNRAS.491.2688V}
{Vida}, D., {Gural}, P.~S., {Brown}, P.~G., {Campbell-Brown}, M., \& {Wiegert},
  P. 2020, \mnras, 491, 2688, \dodoi{10.1093/mnras/stz3160}

\end{thebibliography}
\bibliographystyle{aasjournal}

\end{document}